\documentclass[showpacs,floatfix,twocolumn,prl]{revtex4}
\usepackage{graphicx,color}
\usepackage{amssymb}

\newcommand*{\mcco}{Mg$_{1-x}$Cu$_x$Cr$_2$O$_4$}
\newcommand*{\cco}{CuCr$_2$O$_4$}
\newcommand*{\mco}{MgCr$_2$O$_4$}

\begin{document}

\title{Total scattering descriptions of local and cooperative
distortions in the oxide spinel Mg$_{1-x}$Cu$_x$Cr$_2$O$_4$ with dilute Jahn-Teller ions}

\author{Daniel P. Shoemaker}\email{dshoe@mrl.ucsb.edu}
\author{Ram Seshadri}\email{seshadri@mrl.ucsb.edu}
\affiliation{Materials Department and Materials Research Laboratory,
University of California, Santa Barbara, CA, 93106, USA}

     
\begin{abstract}
The normal spinel oxide MgCr$_2$O$_4$ is cubic at room temperature while the 
normal spinel CuCr$_2$O$_4$ is tetragonal as a consequence of the Jahn-Teller activity 
of Cu$^{2+}$ on the tetrahedral sites. Despite different
end-member structures, a complete solid solution of Mg$_{1-x}$Cu$_x$Cr$_2$O$_4$ can be prepared
with compounds of composition $x$ = 0.43 displaying a first-order phase
transition at room 
temperature. Reverse Monte Carlo analysis of total neutron scattering on data 
acquired between 300 K and 15 K on samples with $x$ = 0.10, 0.20, and 0.43 
provides unbiased local and average structure descriptions of the samples,
including an understanding of the transition from local Jahn-Teller distortions
in the cubic phase to cooperative distortions that result in a tetragonal 
structure. 
Distributions of continuous symmetry 
measures help to understand and distinguish distorted and undistorted 
coordination around the tetrahedral site in the solid solutions.
Magnetic exchange bias is observed in field-cooled hysteresis loops
of samples with dilute Cu$^{2+}$ concentration and in samples
with tetragonal--cubic phase coexistence around 300 K.

\end{abstract}

\maketitle

\section{Introduction}

\begin{figure}
\centering\includegraphics[width=0.8\columnwidth]{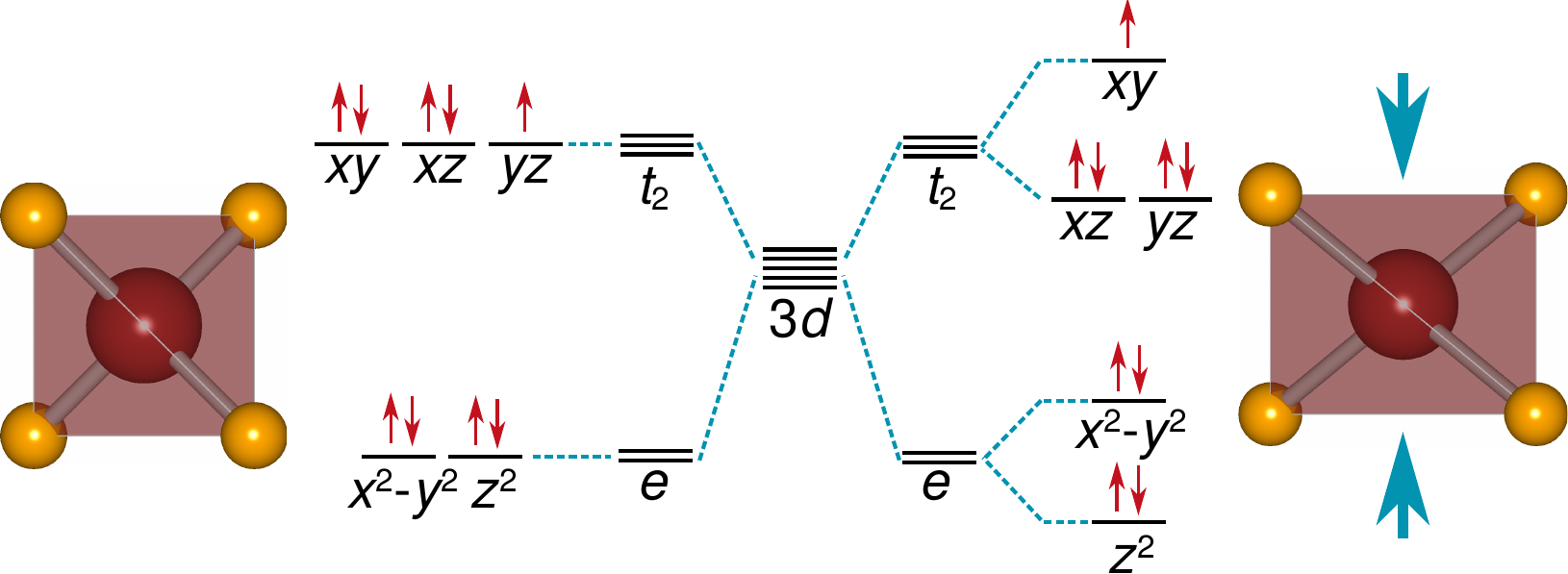}\\
\caption{(Color online) Crystal-field splitting of Cu$^{2+}$ with a $3d^9$ 
electron configuration in an ideal tetrahedron (left) results in 
degeneracy in the $t_2$ orbitals. Jahn-Teller compression of the tetrahedron 
lifts the degeneracy (right), and results in a singly-occupied $xy$ orbital.}
\label{fig:crystalfield}
\end{figure}

The propensity of octahedral Cu$^{2+}$ ions in oxide structures to display
Jahn-Teller (JT) distortions is intimately linked to magnetism and
superconductivity in systems derived from
La$_2$CuO$_4$.\,\cite{bednorz_possible_1986,pavarini_band_2001}
While less common, tetrahedral Cu$^{2+}$ on the $A$ site of oxide spinels
can also display JT activity. This distortion lowers symmetry by
compressing the tetrahedron and thereby breaking the degeneracy
of the partially-occupied $t_2$ energy levels.\,\cite{jahn_stability_1937,gerloch_sense_1981}
The crystal field splittings of the ideal and distorted tetrahedra are
shown schematically in
Fig.\,\ref{fig:crystalfield}.

Jahn-Teller distortions themselves are an intriguing theme in functional
materials because they enable interplay between electronic and structural degrees
of freedom. They have been most widely studied in the manganites,
often derivatives of perovskite
LaMnO$_3$.\,\cite{dagotto_nanoscale_2003,qiu_orbital_2005,sartbaeva_quadrupolar_2007}
In these compounds, Mn$^{3+}$ has four $3d$ electrons
with a singly-occupied pair of $e_g$ states in an octahedral crystal
field. It is well established that elongation of the octahedron breaks
the degeneracy and lowers the energy of the
system.\,\cite{goodenough_theoretical_1955,englman_cooperative_1970}
The percolative nature of
orbital ordering arising from cooperative JT distortion is believed to play
a central role in colossal magnetoresistive behavior.\,\cite{uehara_percolative_1999}

In spinels, collective JT distortions on the $A$ or $B$ sites result in
a reduction in symmetry from cubic $Fd\overline{3}m$ to tetragonal $I4_1/amd$ upon
orbital ordering at the Jahn--Teller transition temperature $T_{JT}$. When only a fraction
of occupied sites are JT-active, cation clustering can lead to
 endotaxial coexistence of tetragonal (distorted) and cubic
phases,\cite{yeo_solid_2006,zhang_coercivity_2007,park_highly_2008}
with strain-driven checkerboard patterns
first studied in phase-separated CoPt
alloys.\,\cite{leroux_electron_1991,le_bouar_origin_1998,fratzl_modeling_1999}
These Mn$^{3+}$-driven JT distortions are
a product of unpaired $3d$ electrons, so self-assembled nanostructured
magnetic films are under development.\,\cite{macmanus-driscoll_self_2010}

Few studies have examined the precise JT tendency of Cu$^{2+}$ on the spinel $A$
site. The effect of Cu$^{2+}$ occupancy on $T_{JT}$ and the
electronic or magnetic properties remains sparesly investigated.
\,\cite{kino_crystal_1966,yan_powder_2007,yan_glassy_2008}
 We
expect there should be key differences between JT activity on the $A$ and
$B$ sites of spinel. While $B$O$_6$ octahedra are edge-sharing and form
a pyrochlore sublattice, the $A$O$_4$ tetrahedra are isolated from each other
in a diamond sublattice. The increased distance between $A$ cations should
hinder their cooperative behavior.

In this work, the structural effects of Cu$^{2+}$ concentration $x$ in the 
spinel solid solution \mcco\ are investigated by real- and reciprocal space structural
probes utilizing total neutron scattering. Discrepancies between the
average and local structural are of particular interest. Specifically,
we probe whether the coordinations of JT-active Cu$^{2+}$ and
JT-inactive Mg$^{2+}$ differ. Traditional Bragg diffraction analysis
fails to resolve these differences because atoms on the same crystallographic
site (here Cu and Mg on the spinel $A$ site) are required to have identical
surroundings. Electron paramagnetic
resonance \cite{waplak_jahnteller_2002} and x-ray absorption \cite{noh_jahn-teller_2006}
offer a confirmation that dilute
Jahn-Teller cations create local distortions, but these spectroscopic
techniques do not yield any detailed structural information.
We utilize the pair distribution function (PDF) because it 
provides a real-space description of the structure with distinct Cu--O
and Mg--O distances.

Our previous PDF study of the spinel CuMn$_2$O$_4$ encountered many of the 
complications that make \mcco\ a difficult crystal structure
to describe.\,\cite{shoemaker_unraveling_2009}  In that study, Cu and Mn
are present with mixed valence on both $A$ and $B$ positions,
and the surrounding oxygen polyhedra are cation-dependent. We found that
CuO$_4$ tetrahedra are significantly more distorted (as judged by bond
angles) than MnO$_4$ tetrahedra, and Cu avoids
the tendency for JT distortion by disproportionating to Cu$^+$/Cu$^{3+}$.
The situation for \mcco\ should be less complex: no valence mixing is present,
and Cu/Mg substitution is confined to the $A$ site. The effect of central
cation on $M$O$_4$ distortion is more isolated.

We employ large box modeling via reverse Monte Carlo simulations as method
of retrieving possible signatures of cation-dependent coordinations from the PDF.
\,\cite{mcgreevy_rmc_1995,mcgreevy_reverse_2001}
Supercells with thousands of atoms can be routinely simulated with modest 
computational requirements, and the large sample size
provides element-specific information due to the presence of many
discrete atoms of each type.
A crucial aspect
of RMC simulations is determining straightforward metrics that describe
how local crystalline structure differs from the average. The
tendency for distortion of individual polyhedra has been
characterized by analyzing bond lengths,\cite{shoemaker_unraveling_2009}
bond angles,\,\cite{shoemaker_unraveling_2009,norberg_local_2009,cambon_effect_2009}
and geometric analysis.\,\cite{wells_reverse_2004,goodwin_ferroelectric_2007}

In this study, tetrahedral JT distortion is gauged using continuous symmetry
measures (CSM).\,\cite{zabrodsky_continuous_1992,pinsky_continuous_1998}
 The particular strength of the 
CSM method is its ability to compare the symmetry of imperfect polyhedra,
regardless of their size or orientation in space.
\,\cite{keinan_studies_2001,ok_distortions_2006}
Extraction of CSM information from RMC simulation was recently performed
as a test of structural rigidity in Bi$_2$Ti$_2$O$_7$ \cite{shoemaker_atomic_2010}
and is employed here to compare the
symmetry of CuO$_4$, MgO$_4$, and $AA_4$ tetrahedra, thereby describing the preference
for JT activity as a function of $x$ and temperature.

\section{Methods}

Powders of \mcco\ compounds were prepared by dissolving stoichiometric amounts
of Cu(NO$_3$)$_2\cdot$2.5H$_2$O, Mg(NO$_3$)$_2\cdot$6H$_2$O, and 
Cr(NO$_3$)$_3\cdot$9H$_2$O in water, followed by boiling to evaporate the solvent 
until a brown mass was formed, which was then ground and calcined
in air at between 700$^\circ$C or 1000$^\circ$C for 10 hours, then cooled
at 10$^\circ$C/min. 
Laboratory X-ray diffraction patterns were acquired using 
Cu-$K_\alpha$ radiation on a Philips X'Pert diffractometer at room temperature 
and a Bruker D8 diffractometer with an Anton Parr high-temperature stage.
Magnetic properties were measured using a Quantum Design MPMS 5XL SQUID 
magnetometer. Time-of-flight (TOF) neutron scattering was performed on the 
NPDF instrument at Los Alamos National Laboratory.  Rietveld refinements were
performed using the \textsc{XND} code\cite{berar_xnd_1998} for 
X-ray data and the \textsc{GSAS-EXPGUI} suite\cite{larson_general_2000} 
for the TOF data. The PDF was extracted using the \textsc{PDFGetN} 
program\cite{peterson_pdfgetn_2000} with $Q_{max} = 35$\,\AA$^{-1}$
and least-squares refinement of the PDF was performed using the 
\textsc{PDFgui} frontend for \textsc{PDFFit2}.\cite{farrow_pdffit2_2007}
Crystal structures were visualized with 
\textsc{AtomEye}\cite{li_atomeye_2003} and 
\textsc{VESTA}.\,\cite{momma_vesta_2008}

Reverse Monte Carlo simulations
were run using \textsc{RMCProfile} version 6 \cite{tucker_rmcprofile_2007} on
$7 \times 7 \times 7$ cubic or $10 \times 10 \times 7$ tetragonal spinel
supercells with 19208 or 19600 atoms, respectively. A hard-sphere 
repulsion was applied to prevent $M$--O bond distances shorter than the
first peak of the PDF, but no clustering was observed at the cutoff distances.
No preference for cation clustering was found, so configurations with randomized
Cu and Mg occupancy were used. Simulations were performed as serial
jobs on the HP Opteron QSR cluster at the California NanoSystems Institute.

Bond valence sums (BVS) were extracted from atoms in the
supercell in the same manner described in our previous work
on CuMn$_2$O$_4$,\cite{shoemaker_unraveling_2009}
using the $R_0$ values of Brese and O'Keeffe.\cite{brese_bond_1991}
CSM for $A$O$_4$ tetrahedra were calculated using a distance measure
program provided by M. Pinsky and D. Avnir.

\section{Results and Discussion}

\subsection{Average structure \emph{via} reciprocal-space analysis}

The compounds \mco\ and \cco\ both belong to the $AB_2$O$_4$ spinel family of
structures with
the $A$ cations, Mg$^{2+}$ and Cu$^{2+}$, tetrahedrally coordinated by oxygen, while 
Cr$^{3+}$ lies on the octahedral $B$ site. The [Ar]$3d^3$ electron configuration of 
Cr$^{3+}$ is very stable because each of the $t_{2g}$ energy levels is singly occupied,
so there is no tendency of site mixing or mixed
valence.\,\cite{miller_distribution_1959} Varying
$x$  in \mcco\ therefore does not disturb the $B$ sublattice composition, but changes
in the interpenetrating $A$ sublattice may lead to chemical pressure which
will influence its size and shape.

\begin{figure}
\centering\includegraphics[width=0.8\columnwidth]{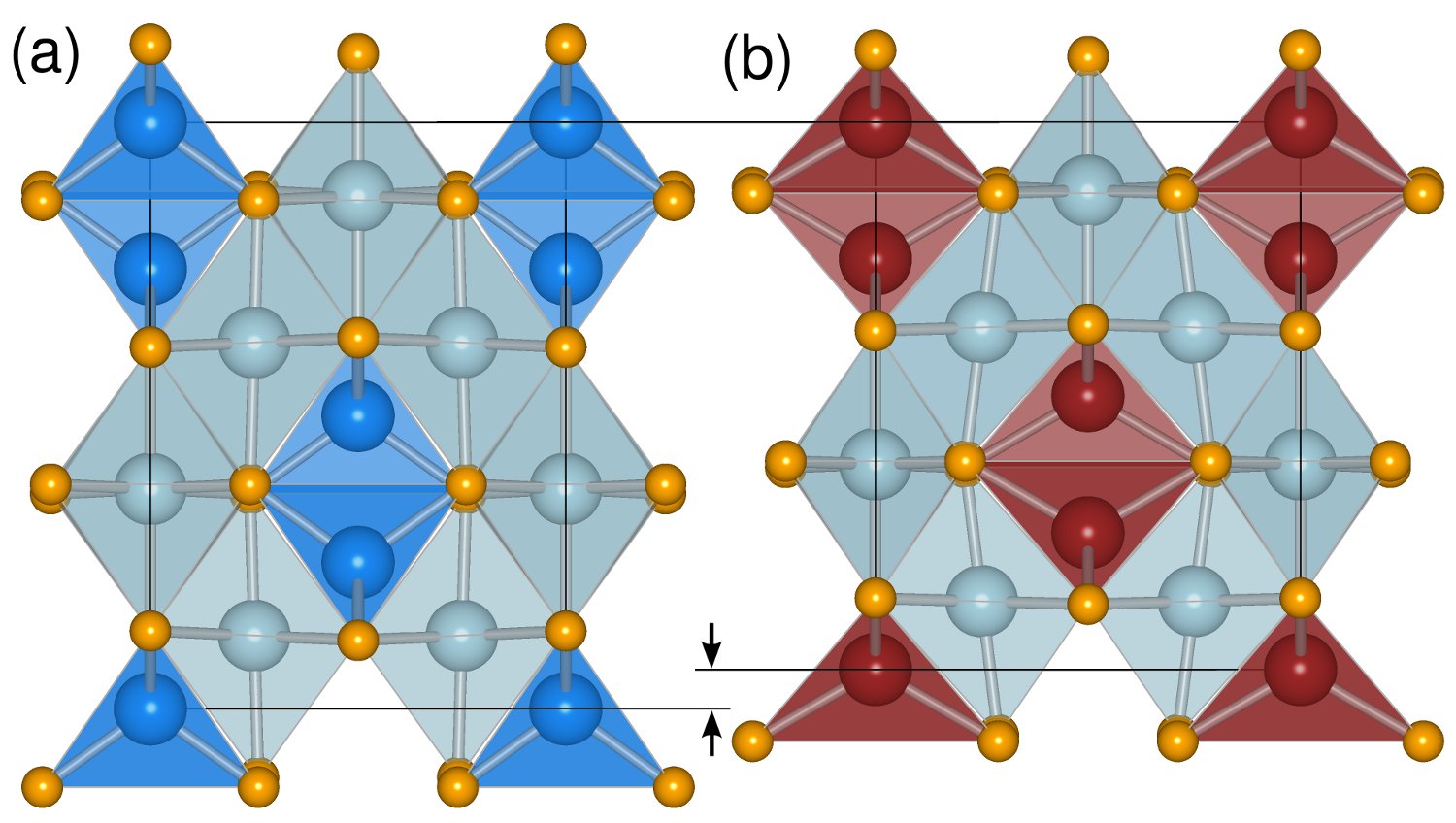}\\
\caption{(Color online)
%The spinel unit cell of \mco\ is shown in (a). Below,
To scale, the pseudotetragonal
cells of (a) cubic \mco\ and (b) Jahn-Teller distorted tetragonal \cco\ are viewed
along the $a$ axis of the $I4_1/amd$ cell. Contraction
in the $c$ direction is evident due to JT distortion in \cco.
CrO$_6$ octahedra are light blue,
while MgO$_4$ and CuO$_4$ tetrahedra are dark blue and red, respectively. 
}
\label{fig:unitcells}
\end{figure}

On the tetrahedral $A$ site, Mg$^{2+}$ and Cu$^{2+}$ have effectively identical ionic
radii. Both are 0.57 \AA\ as given by Shannon,\cite{shannon_revised_1976}
but their electron configurations are distinctly
different. Mg$^{2+}$ has the [Ne] configuration and no $d$ electrons.
Cu$^{2+}$ has [Ar]$3d^9$ and only two of the three $t_2$ 
energy levels are fully occupied in tetrahedral coordination (Fig.\,\ref{fig:crystalfield}).
This degeneracy causes
a JT distortion, manifested by a flattening of the tetrahedron.
Bond lengths are preserved, but bond angles are no longer equivalent 
at 109.5$^\circ$.\,\cite{dollase_spinels_1997,greenwood_chemistry_1997,day_theoretical_2009}
The contrasting behavior of Mg$^{2+}$ and Cu$^{2+}$ in tetrahedral coordination
is evident when the \mco\ and \cco\ structures are compared in Fig.\,\ref{fig:unitcells}.
\mco\ forms in the cubic space group $Fd\overline{3}m$ with ideal MgO$_4$ 
tetrahedra.\cite{verwey_physical_1947}
\cco\ undergoes flattening in the $c$ direction and forms
in the tetrahedral space group $I4_1/amd$, which is the same space group as
JT distorted Mn$_3$O$_4$ (and other $A$Mn$_2$O$_4$) or
NiCr$_2$O$_4$.\cite{kennedy_role_2008, dollase_spinels_1997}
The unit cells are shown to scale in 
Fig.\,\ref{fig:unitcells} to highlight their difference in dimensions.

\begin{figure}
\centering\includegraphics[width=0.9\columnwidth]{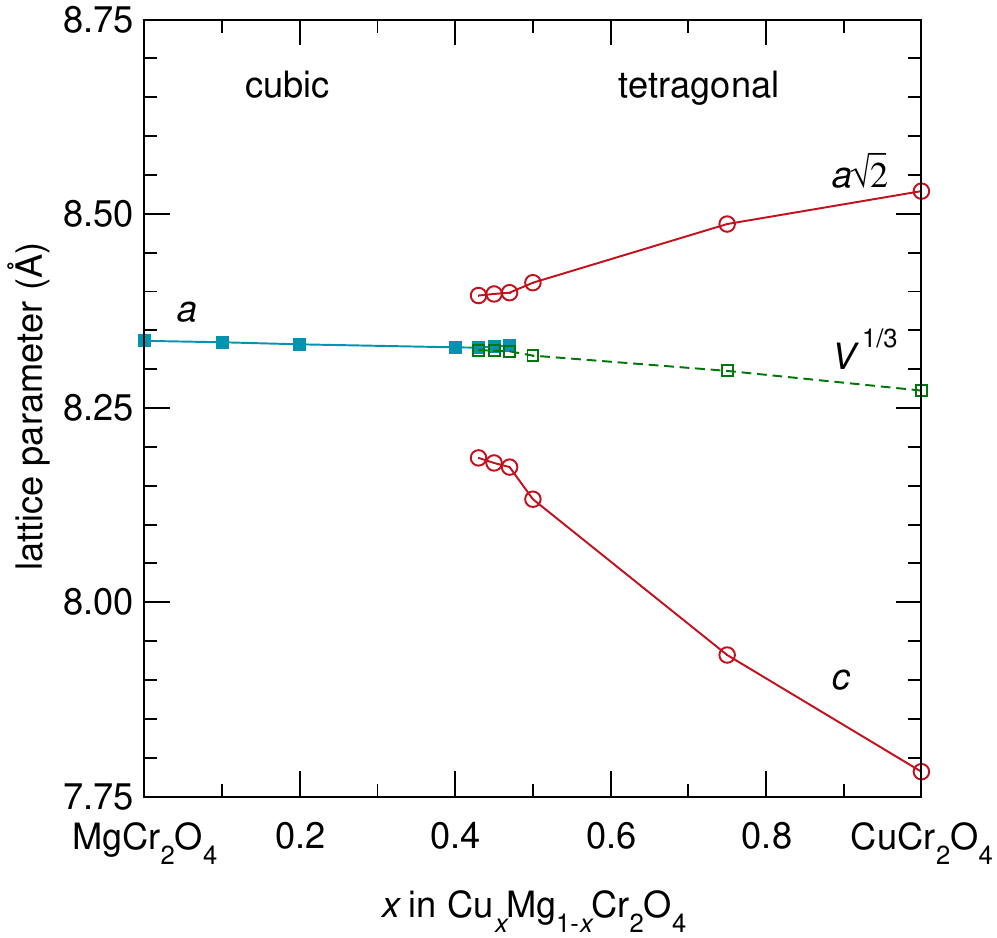}\\
\caption{(Color online) Lattice parameters
of the \mcco\ solid solution obtained by
Rietveld refinement of laboratory X-ray powder diffraction at room temperature.
Coexistence of the cubic and tetragonal phases occurs for 0.43 $\leq x \leq$ 0.47.
$V^{1/3}$ is shown for tetragonal phases. Error bars are smaller than the symbols
for all points.
}
\label{fig:latparams}
\end{figure}

Alloying $x$ from 0 to 1 in \mcco\ produces a transition from a
cubic to tetragonal spinel. The lattice parameters
obtained from Rietveld refinements to room-temperature XRD patterns
shown in Fig.\,\ref{fig:latparams} reveal
that for $x < 0.43$ a cubic spinel is formed with a gradually
decreasing lattice parameter $a$. When $x = 0.43$ the tetragonal
phase appears with $c/a = 0.975$ and a small region of coexistence persists
for $0.43 \leq x \leq 0.47$, above which the cubic phase disappears.
The compound becomes increasingly
tetragonal as the end member \cco\ is approached, with
$c/a = 0.912$ when $x = 1$. The pseudocubic
cell volume contracts from 579.4 \AA$^3$ for \mco\ to 566.1 \AA$^3$ for 
\cco, which is a 2.3\% decrease.

\begin{figure}
\centering\includegraphics[width=0.9\columnwidth]{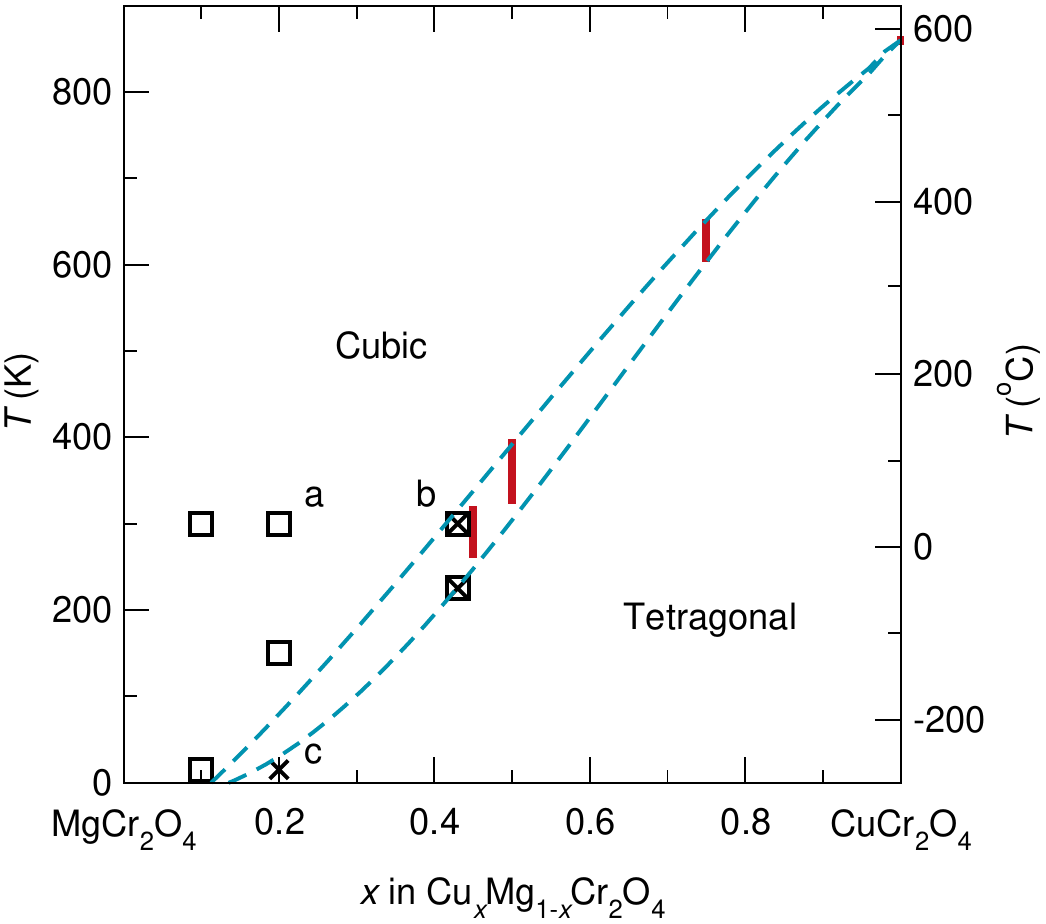}\\
\caption{(Color online) Phase diagram of the
\mcco\ system as determined by Rietveld
refinement. Points denote neutron refinements to cubic ($\square$) and tetragonal ($\times$)
phases. Bars represent coexistence regions from high-temperature X-ray diffraction.
Letters (a,b,c) correspond to the Rietveld refinements shown in Fig.\,\ref{fig:rietvelds}.
}
\label{fig:phasediagram}
\end{figure}

\begin{figure}
\centering\includegraphics[width=0.8\columnwidth]{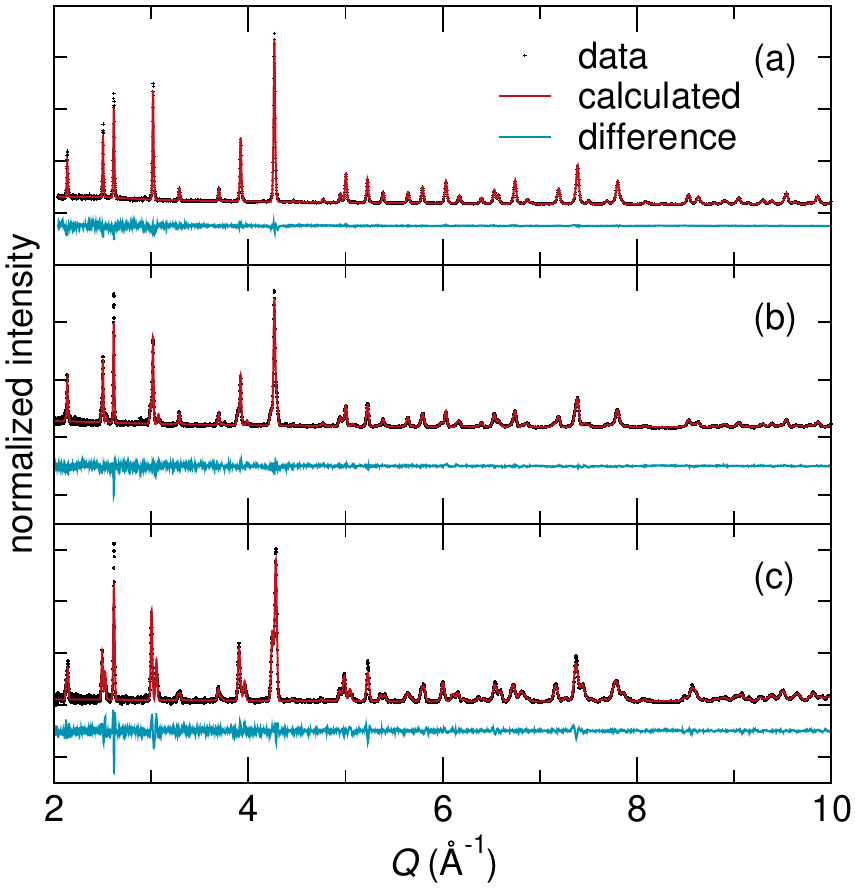}\\
\caption{(Color online) Time-of-flight neutron scattering
Rietveld refinements
of \mcco\ with (a) $x$ = 0.20 at 300 K [cubic], (b) $x$ = 0.43 at 300 K [cubic
and tetragonal], and (c) $x$ = 0.20 at 15 K [tetragonal].}
\label{fig:rietvelds}
\end{figure}

We extend our Rietveld analysis using high-temperature XRD and low-temperature
TOF neutron scattering to produce an approximate phase diagram
of the pseudobinary system \mco--\cco, shown in Fig.\,\ref{fig:phasediagram}.
The $T_{JT}$ between cubic and tetragonal spinels has a nearly linear relationship
on $x$. There is some phase coexistence determined from HTXRD
denoted by the bars on the graph. We use the end-member \cco\ transition
of $T_{JT}$ = 590$^\circ$C from the
literature.\,\cite{tovar_structural_2006,kennedy_role_2008}
The JT transition temperature steadily decreases with decreasing $x$ so that
the tetragonal phase occurs for $x$ = 0.20 but not for $x$ = 0.10 at $T$ = 15~K (the lowest
temperature measured).
Rietveld refinements to neutron scattering data are shown in Fig.\,\ref{fig:rietvelds}
for three different points on the phase diagram, representing (a) cubic, (b)
mixed, and (c) tetragonal phases at $x$ = 0.20 and $T$ = 300 K, $x$ = 0.43 and
$T$ = 300~K, and $x$ = 0.20 and $T$ = 15 K, respectively. In all cases,
the fits are excellent. There is, however, some unfit intensity between
split tetragonal peaks in (b) and (c), indicative of a more complex crystal
structure than the two-phase Rietveld model would suggest.

Not shown on our phase diagram is the tetragonal distortion around 12 K 
in \mco.\cite{ehrenberg_tetragonal_2002,klemme_low-temperature_2007,ortega-san-martin_low_2008}
A transition from $Fd\overline{3}m$ to $I4_1/amd$ occurs
to remove the geometric frustration of Cr$^{3+}$ spins arranged in a pyrochlore
sublattice.
Addition of magnetic Cu$^{2+}$ cations in the $A$ sites should relieve this
frustration due to strong $A$-$B$ interactions,
\cite{zhang_local_2006,yan_glassy_2008} so we do not expect
the spin-driven distortion to play a role for $x > 0$. 
The sample with $x$ = 0.10 is cubic at 15 K.

The presence of a tetragonal phase for low Cu$^{2+}$ content is surprising.
At $x$ = 0.20, for instance, only one in five $A$ sites has a JT active cation.
The $A$ cations are arranged in a diamond sublattice and 
each has four nearest $A$ neighbors. The tetrahedra do not
share edges or corners with each other, with the shortest exchange
pathway being $A$--O--Cr--O--$A$. Given a random
cation distribution, the probability of one Cu$^{2+}$ having all JT-inactive
Mg$^{2+}$ neighbors is $(\frac{4}{5})^4 = 41.0\%$. The probability of having
only one Cu$^{2+}$ neighbor is $4(\frac{1}{5})(\frac{4}{5})^3 = 41.0\%$,
and the probability of having two Cu$^{2+}$ neighbors falls to
$6(\frac{1}{5})^2(\frac{4}{5})^2 = 15.4\%$. Thus 82.0\% of Cu$^{2+}$
cations have zero or only one JT-active nearest neighbor, but they
still produce orbital ordering with long-range periodicity. The trend
of $T_{JT}$ versus $x$ is roughly linear in Fig.\,\ref{fig:phasediagram},
with no apparent jump
at the percolation threshold of the diamond-type $A$
sublattice.\,\cite{marck_percolation_1997,shoemaker_intrinsic_2009}

The critical concentration of Cu$^{2+}$ needed to drive a cooperative
JT distortion in \mcco\ at 300 K is $x$ =  0.43. This fraction
increases with $A$-site cation radius in $A$Cr$_2$O$_4$ spinels:
Zn$_{1-x}$Cu$_x$Cr$_2$O$_4$ ($r_{Zn}$ = 0.60 \AA) is reported to have
$x = 0.47$ \cite{reinen_local_1988} and
$x = 0.58$, \cite{yan_glassy_2008} while
Cd$_{1-x}$Cu$_x$Cr$_2$O$_4$ ($r_{Cd}$ = 0.78 \AA) has $x = 0.64$.\,\cite{yan_powder_2007}
This could be due to increased distance between $A$-site cations, or
a loosening of the structure (thus weakening of strain field
produced by a JT distortion).

The critical concentration in 
Zn$_{1-x}$Ni$_x$Cr$_2$O$_4$,\,\cite{kino_cooperative_1972}
where Ni$^{2+}$ drives JT disortion, is around $x \sim 1$ at room temperature.
For Cr$^{3+}$ on the spinel $B$ site, less Cu$^{2+}$ is needed to drive a
cooperative distortion than Ni$^{2+}$. The $3d^8$ configuration of Ni$^{2+}$
has only one
unpaired $t_2$ electron, rather than the two of Cu$^{2+}$. The result
is a smaller energy gain after breaking degeneracy and
elongation (rather than contraction) of the $c$ axis.\,\cite{gerloch_sense_1981}

Comparison with JT tendency of Mn$^{3+}$ on the spinel $B$ site is less direct.
For example, the solid solution Zn[Fe$_{1-x}$Mn$_x$]$_2$O$_4$ has a critical
concentration of about $x$ = 0.3, \cite{okeeffe_cation_1961} while Mn[Cr$_{1-x}$Mn$_x$]$_2$O$_4$
has $x$ = 0.4.\cite{holba_tetragonal_1975} This would seem to indicate a stronger JT tendency, in part
due to closer $B$--$B$ distances and edge sharing between octahedra. However,
for Zn$_{x/2}$Ge$_{1-x/2}$[Co$_{1-x}$Mn$_x$]$_2$O$_4$, Wickham reports
$x$ = 0.65, \cite{wickham_crystallographic_1958} and Bhandage reports $x$ = 0.70 for
Zn$_{x/2}$Mn$_{1-x/2}$[Ni$_{1-x}$Mn$_x$]$_2$O$_4$.\cite{bhandage_magnetic_1978}
The wide spread in critical concentrations of Mn$^{3+}$ can be explained by the 
differences in JT splitting energies found by X-ray absorption spectroscopy
on $A$Mn$_2$O$_4$ spinels by Noh, \emph{et al}.\,\cite{noh_jahn-teller_2006} In essence, the 
energy drop from JT distortion around Mn$^{3+}$ on the spinel $B$ site
is very senstive to changes in chemical pressure.

We obtain the \emph{cooperative} behavior of the averaged lattice using Rietveld
refinement. We do not necessarily resolve the distinct
cation coordinations of Mg$^{2+}$ and Cu$^{2+}$ if they are different on
the local, atomic length scale. Two views of the JT transition can be proposed:
in the case of a sharp crossover, as would be implied by how 
Rietveld analysis is performed, all $A$O$_4$ tetrahedra are equivalent
whether they contain Mg$^{2+}$ or Cu$^{2+}$, and upon increasing $x$ they 
abruptly transform from ideal tetrahedra in the cubic spinel to flattened tetrahedra
in the cooperatively JT distorted spinel. In the second case, 
the CuO$_4$ tetrahedra are \emph{always} locally JT distorted (even for values of
$x$ where the spinel is cubic) but the crossover at $x$ = 0.43 at
room temperature represents
the point where they cooperatively order and the JT distortions
percolate through the long-range structure.

\subsection{Local structure \emph{via} real-space analysis}

The average structure model of \mcco\ from Rietveld refinement
indicates that the compounds exist
as single phases, either cubic or tetragonal, apart from
the two-phase coexistence region around $T_{JT}$. When modeled using a
single unit cell, Mg$^{2+}$ and Cu$^{2+}$ are required to share the same crystallographic
site and their surroundings are necessarily identical. 
This model often inadequately describes the true structure of compounds
where energy-lowering changes in cation coordination
are known to persist above the average structural 
transition temperature as in perovskite manganites and cobaltites.
\cite{qiu_orbital_2005, bozin_understanding_2007,sartbaeva_quadrupolar_2007,louca_dynamical_2003}
The PDF has emerged as a key tool for measuring
these local distortions that do not possess long-range order. Because
the PDF is a weighted histogram of all atom-atom distances in the sample,
it is sensitive to the distinct bond distances that are produced by
dissimilar coordination of multiple chemical
species on the same site.\,\cite{shoemaker_unraveling_2009} We investigate whether the PDF shows any
signature of distinct Cu$^{2+}$ and Mg$^{2+}$ coordination.

\begin{figure}
\centering\includegraphics[width=0.9\columnwidth]{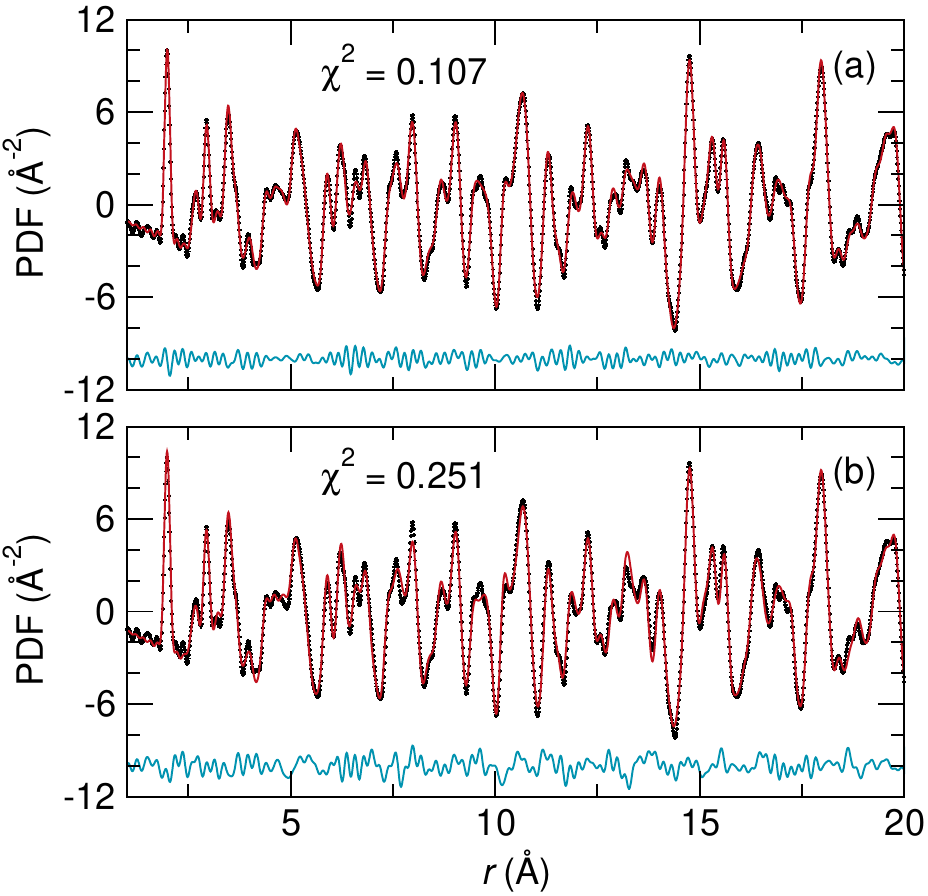}\\
\caption{(Color online) Least-squares fits to the PDF for $x$ = 0.20 at
300~K for the (a) cubic spinel phase with fractional-occupancy obtained from
Rietveld refinement, and (b) an 80-20 fit to the end members \mco\ and \cco.
}
\label{fig:pdfgui-cmco20}
\end{figure}

Least-squares PDF refinements can be performed using the average structure
unit cells from Rietveld refinement as a starting point. Fits to the
$x$ = 0.20 data at 300~K are shown in Fig.\,\ref{fig:pdfgui-cmco20}. Panel
(a) shows the fit to a Rietveld-refined cubic unit cell with split 0.20/0.80
occupancy of Cu$^{2+}$ and Mg$^{2+}$ on
the same crystallographic site. In (b),  
we fit using a 0.20/0.80 linear combination of the \mco\ and \cco\
end members with lattice parameters allowed to refine.
The fit is good despite the use
of a tetragonal unit cell to model a structure
far above $T_{JT}$, but it does not improve on the fit
using a single cubic unit cell.
We use the value
\begin{equation}
\chi^2 = \sum{}{}{\frac{(\mathrm{PDF}^{obs}-\mathrm{PDF}^{calc})^2}{N}}
\end{equation}
to compare fits, where $N$ is the number of points in the PDF from
the nearest-neighbor cutoff to 20~\AA.
%% Similarities between the two fits
%% imply that it is difficult to discern
%% the contribution of distorted CuO$_4$ tetrahedra for low Cu$^{2+}$ concentrations.
The ability to distinguish Mg and Cu is hindered by their similar neutron
scattering lengths: 5.38 and 7.71~fm, respectively.\,\cite{sears_neutron_1992}
For $x$ = 0.20, least-squares PDF fits do not definitively prove that there
are distinct (or identical) coordination environments for Mg$^{2+}$
and Cu$^{2+}$.

\begin{figure}
\centering\includegraphics[width=0.9\columnwidth]{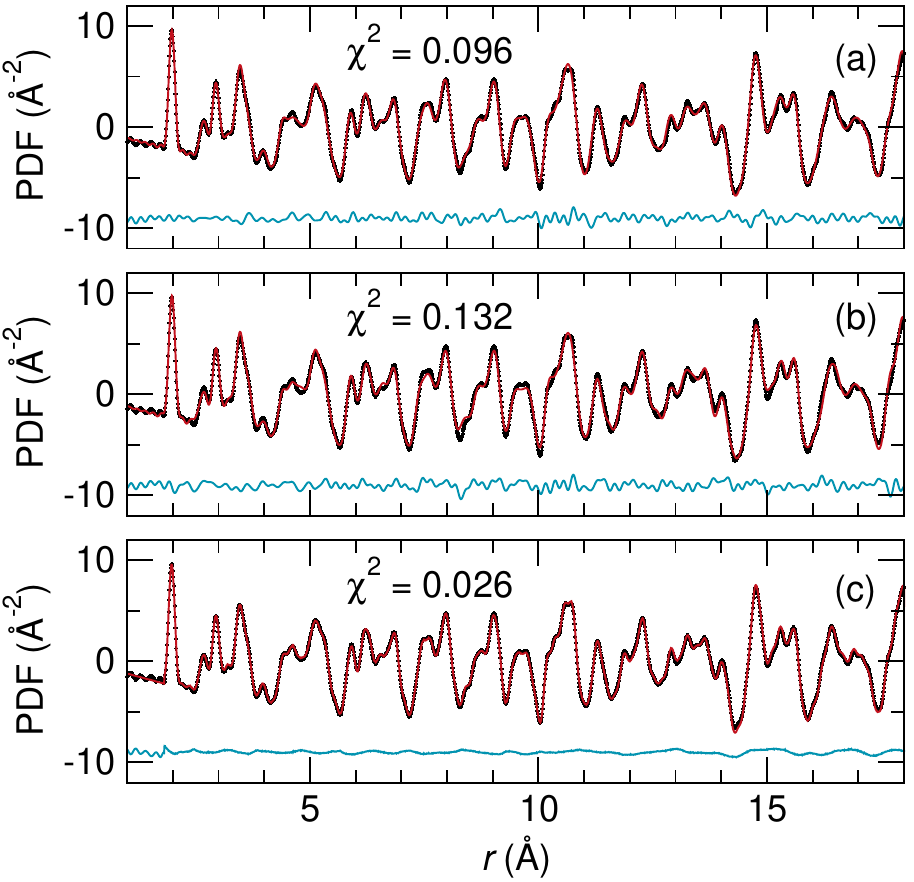}\\
\caption{(Color online) Fits to the PDF for $x$ = 0.43 at 300~K for (a)
least-squares refinement of the two fractional-occupancy phases found
from Rietveld refinement, (b) least-squares refinement using
the end members \mco\ and \cco, and (c) the fit after RMC simulation.
}
\label{fig:pdfgui-cmco43}
\end{figure}

Resolution of distinct cation environments is aided when $x$ = 0.43 due
to approximately even concentrations of Cu$^{2+}$ and Mg$^{2+}$.
At 300~K, Rietveld refinement found coexistence of the cubic and tetragonal phases.
Neither phase alone can be used to produce a satisfactory fit to the PDF.
A two-phase fit using the Rietveld refined cells
is shown in Fig.\,\ref{fig:pdfgui-cmco43}(a), and agrees quite well with the 
data. As with the $x$ = 0.20 sample, we also fit the data to a
combination of the end members \mco\ and \cco\ in Fig.\,\ref{fig:pdfgui-cmco43}(b).
Again, the fractional occupancy Rietveld result produces a better fit than the
end members. The least-squares fits indicate that the average structures
produce excellent representations of the local structures, but they
do not definitively show whether the Mg$^{2+}$ and Cu$^{2+}$ coordination
environments are distinct or similar. 

Least-squares fits to the PDF are required to specify how many
distinct $A$O$_4$ environments to allow, much like in a Rietveld refinement.
The $Fd\overline{3}m$ and $I4_1/amd$ unit cells provide only one
$A$O$_4$ environment per phase. There is no way to define cation-dependent
coordination without manually building a lower-symmetry unit cell.
In order to investigate the $A$O$_4$
environment directly, we remove the symmetry constraints of
least-squares PDF analysis and utilize large-box modeling via reverse Monte Carlo
(RMC) simulations. This method has proved to be useful for investigating
atomic structure on the local level, especially in cases where long-range
periodicity is not present, such as SrSnO$_3$,\cite{goodwin_ferroelectric_2007}
Bi$_2$Ti$_2$O$_7$,\cite{shoemaker_atomic_2010} and
$\beta$-cristobalite.\,\cite{tucker_dynamic_2001}

\begin{figure}
\centering\includegraphics[width=0.7\columnwidth]{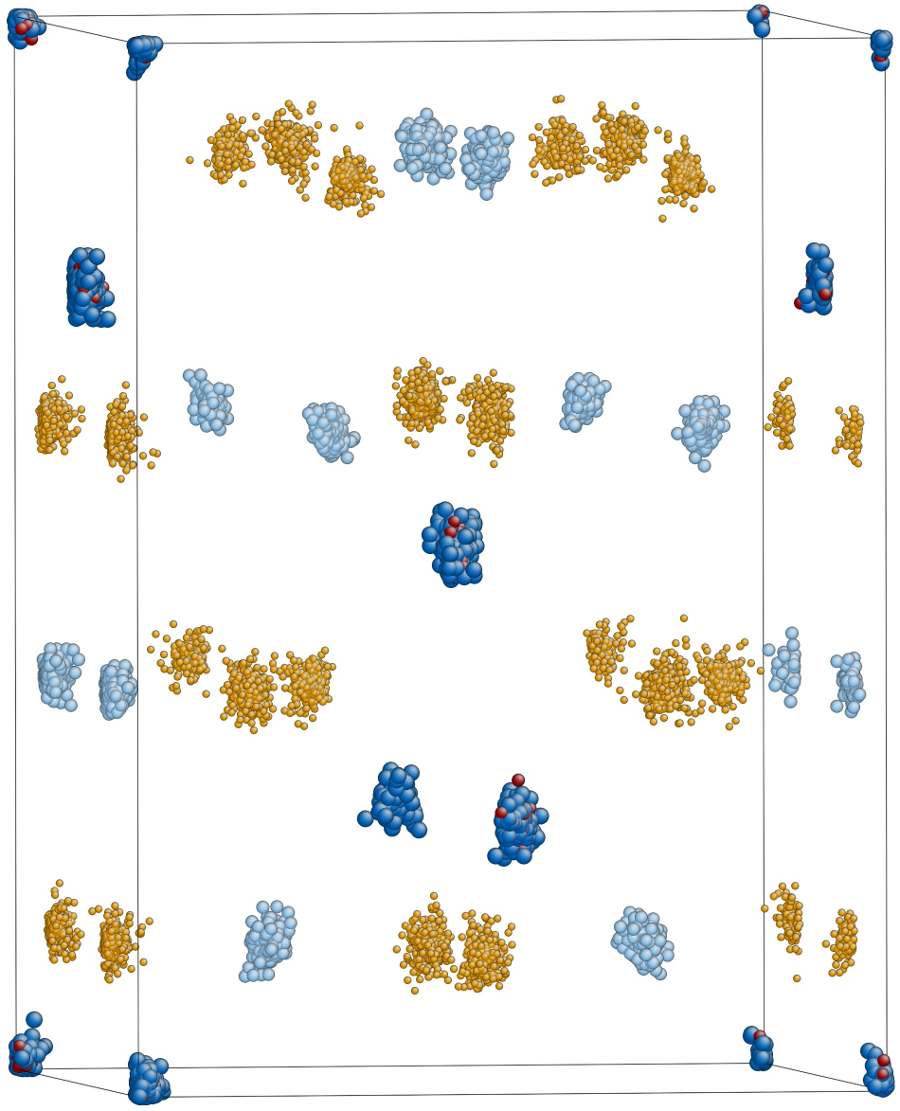}\\
\caption{(Color online) The final $10 \times 10 \times 7$ RMC supercell for 
an $x$ = 0.20 simulation is
folded into a single unit cell to obtain point clouds at each
crystallographic site. Atoms shown are Cu (red), Mg (dark blue), Cr (light blue), 
and O (orange).}
\label{fig:supercell}
\end{figure}

\begin{figure}
\centering\includegraphics[width=0.9\columnwidth]{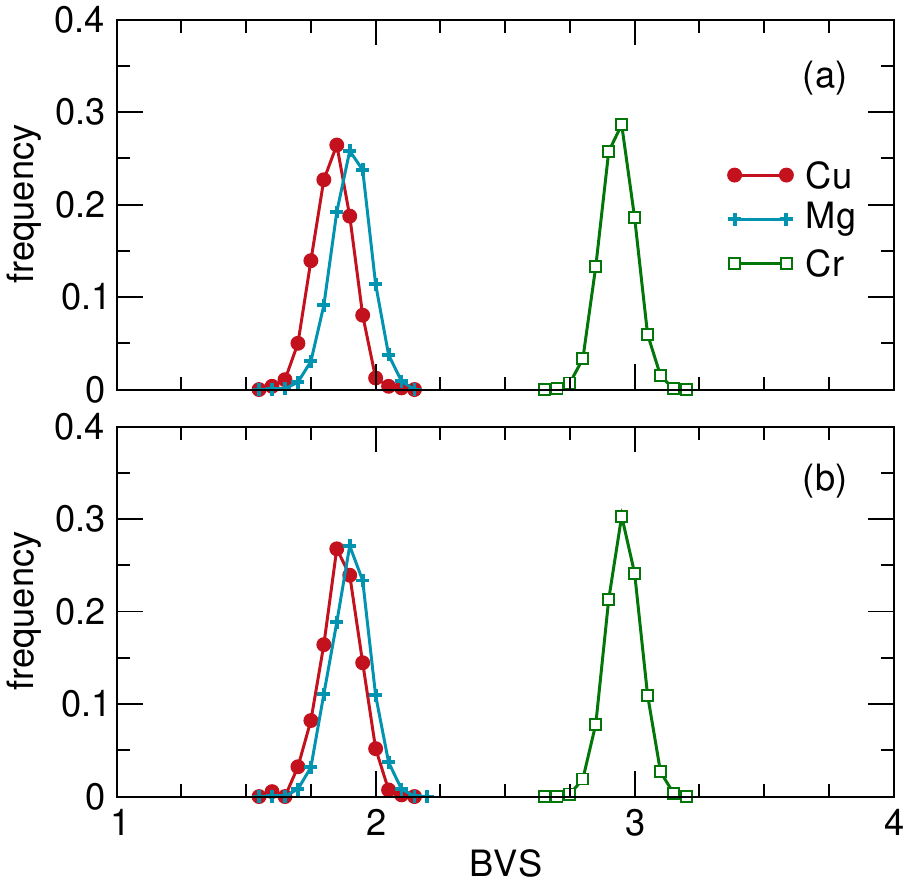}\\
\caption{(Color online) Bond valence sums for Cu and Mg extracted from \mcco\ RMC supercells
$x$ = 0.20 after fits to data at $T$ = 300 K (a) and 15 K (b). Aside from a slight
broadening at 300 K, the distributions are comparable. }
\label{fig:bvs}
\end{figure}

The RMC supercell for an $x$ = 0.20 sample at 15 K is inspected by folding
each of the unit cells back into a single box, shown in
Fig.\,\ref{fig:supercell}, which reveals how each crystallographic site is
decorated with atoms. The supercell contains 560 Cu and 2240 Mg atoms that
are randomly arranged. Because there are a large number of distinct Cu and Mg
atoms, statistical analysis can be used to investigate whether there is any
evidence for the local $A$O$_4$ distortion to depend on the central cation.
Bond valence sum histograms shown in Fig.\,\ref{fig:bvs} show that both the $A$ cations
have valences peaked around the expected value of $A^{2+}$, and the $B$ site
shows only Cr$^{3+}$, which implies that our supercells contain chemically
reasonable bond lengths.

%% One advantage of RMC is the ability to fit to real- and reciprocal-space structures
%% simultaneously. The PDF, structure factor $F(Q)$, and Bragg profile can be used
%% as constraints in the simulation. [but here we are not fitting to k-space data]

No cation dependence of $A$O$_4$ bond distances is obvious from partial
radial distributions $g_{Cu-O}(r)$ and $g_{Mg-O}(r)$. This
is to be expected because Mg$^{2+}$ and Cu$^{2+}$
have the same ionic radius when tetrahedrally coordinated. Neither is there
any apparent distinction between CuO$_4$ and MgO$_4$ tetrahedra based upon
O--$A$--O bond angles, in contrast to our previous study on
CuMn$_2$O$_4$.\cite{shoemaker_unraveling_2009}
%% This can be inherently difficult to resolve because it requires proof that
%% the bond distribution peak is composed of two gaussians rather than one.

\begin{figure}
\centering\includegraphics[width=0.9\columnwidth]{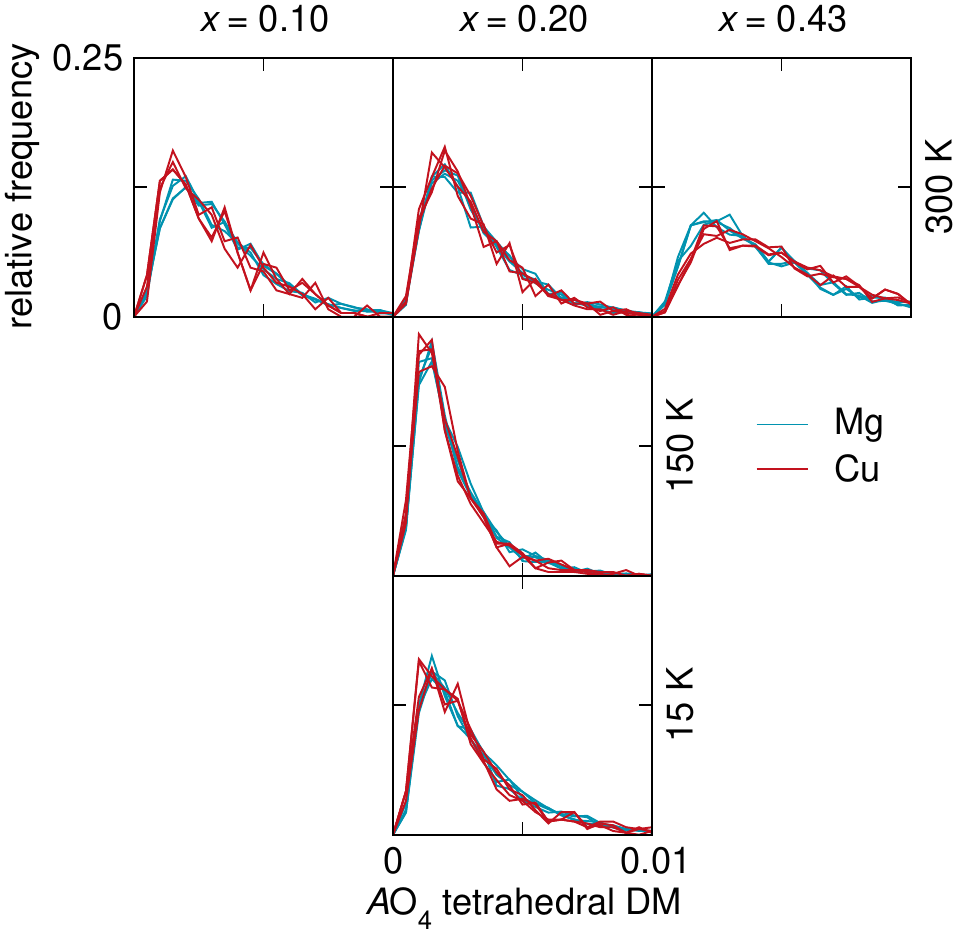}\\
\caption{(Color online) CSM histograms from RMC simulations for three
Cu contents and three temperatures. Each panel contains 8 distributions: one
Cu and one Mg for four independent simulations. Broadening 
across the $T$ = 300~K series indicates increasing tetrahedral distortion
with $x$.
Thermal effects lead to sharpening and shifting toward smaller DM when the
 $x$ = 0.20 sample is cooled to 150 K. This tendency toward
ideal tetrahedra is disrupted by long-range tetragonal distortion upon
further cooling, so the $T$ = 15~K histogram is broad and shifted to higher DM.}
\label{fig:csm-hist}
\end{figure}

We use the continuous symmetry measure (CSM) technique to gauge the tendency
for JT distortion of $A$O$_4$ tetrahedra. The CSM technique provides a distance
measure (DM) of a given tetrahedron that indicates its deviation from ideality.
\,\cite{zabrodsky_continuous_1992,pinsky_continuous_1998}
A perfect tetrahedron has DM = 0, and any distortion increases the value
of DM. In the end member compounds, MgO$_4$ has a DM = 0 while CuO$_4$ has a
DM = 0.0076. Having thousands of distinct tetrahedra in the RMC supercell affords
the opportunity to produce a histogram of DM (Fig.\,\ref{fig:csm-hist})
for all tetrahedra depending on the central cation. Each
panel contains eight lines: four for each cation, resulting from four independent
RMC simulations. The overall shape of each histogram describes
the average distortion (peak center) and the tightness of the DM distribution 
(peak width) at each value of $x$ and $T$. No tetrahedra are present
with exactly DM = 0 because stochastic RMC simulations leave no atomic
positions untouched--even a compound with ideal tetrahedra would
have the shapes subtly distorted.

At 300 K, the spread of distortions
increases for $x$ = 0.43 because the average structure becomes a mixture
of cubic and tetragonal phases. In the series where $x$ = 0.20, the peak sharpens
upon cooling to 150 K, which we attribute to a reduction of thermal vibrations. It also
moves to smaller DM values, indicating a tendency toward a more ideal
$M$O$_4$ environment. This
trend would continue to low temperature in the absence of long-range
JT distortion. Instead,
the peak broadens and shifts to higher DM at 15 K. We attribute both
of these effects to the tetragaonal phase transition. These histograms
provide a view of the average tetrahedral shape, but no distinction between
Mg and Cu is apparent. We find that cumulative distributions offer a 
clearer picture of this dependence.

\begin{figure}
\centering\includegraphics[width=0.9\columnwidth]{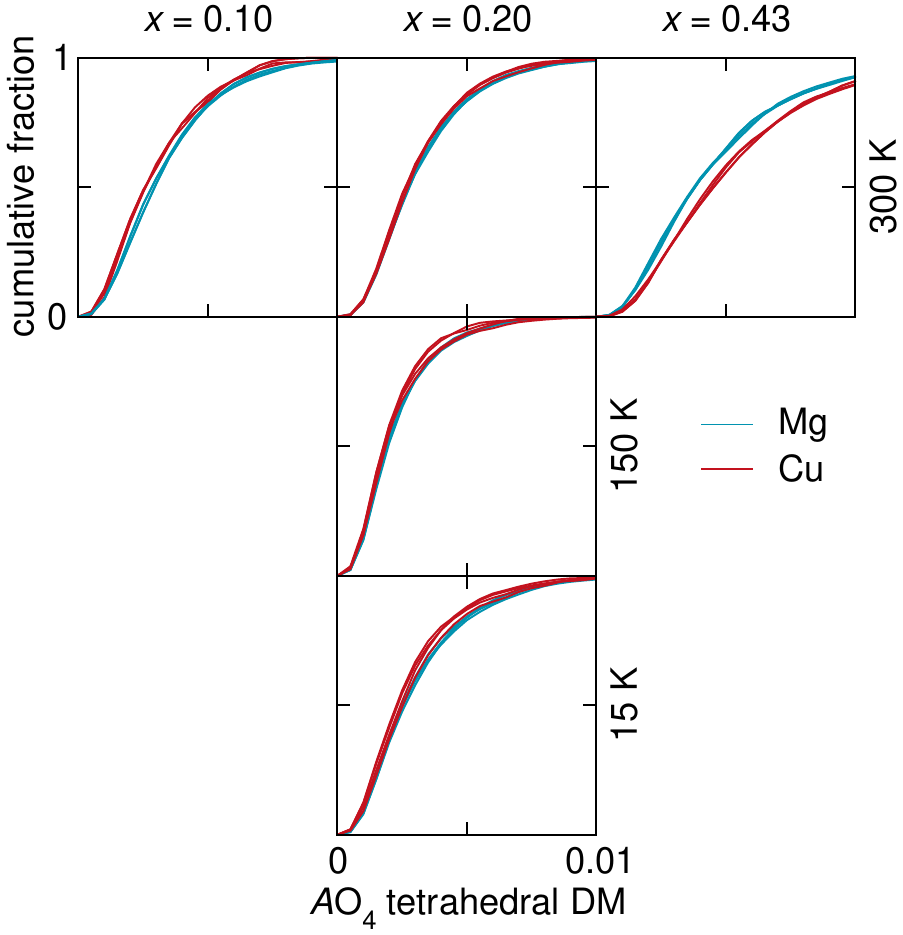}\\
\caption{(Color online) Cumulative CSM distributions are similar
enough from run to run that most lines overlay each other. Only the $x$ = 0.43 run has a
clear distinction between CuO$_4$ and MgO$_4$ tetrahedron shape, with MgO$_4$ distinctly
more ideal than CuO$_4$.}
\label{fig:csm-cumu}
\end{figure}

The cumulative distributions in 
Fig.\,\ref{fig:csm-cumu} do not show separation between the Cu and Mg curves in the $x$ = 0.10
or 0.20 samples.
However, the $x$ = 0.43 sample shows a clear distinction, with Mg tetrahedra
possessing DM that are closer to zero (more ideal) than Cu. This is clear evidence for the
tendency of Cu to undergo JT distortion while Mg remains more symmetric.

Overlapping DM curves for $x$ = 0.10 and 0.20 do not preclude the possibility of
distinct CuO$_4$ and MgO$_4$ tetrahedra in those samples. There may be
insufficient resolution in the PDF to
distinguish the cations due to their similar neutron scattering cross sections
and the relatively low concentration of Cu$^{2+}$. It remains unclear 
whether distinct cation environments are only seen around
$x$ = 0.5 (due to approximately even cation concentrations) and if the distinction
would disappear as $x \rightarrow 1$. Temperature dependence of the CSM when $x$ = 0.43
may also provide some insight into the dynamics of these distortions.
Substituting a JT-inactive $A$ site cation with a different neutron scattering
cross length, such as Mn$^{2+}$ or Co$^{2+}$, may aid contrast with Cu$^{2+}$.

\begin{figure}
\centering\includegraphics[width=0.9\columnwidth]{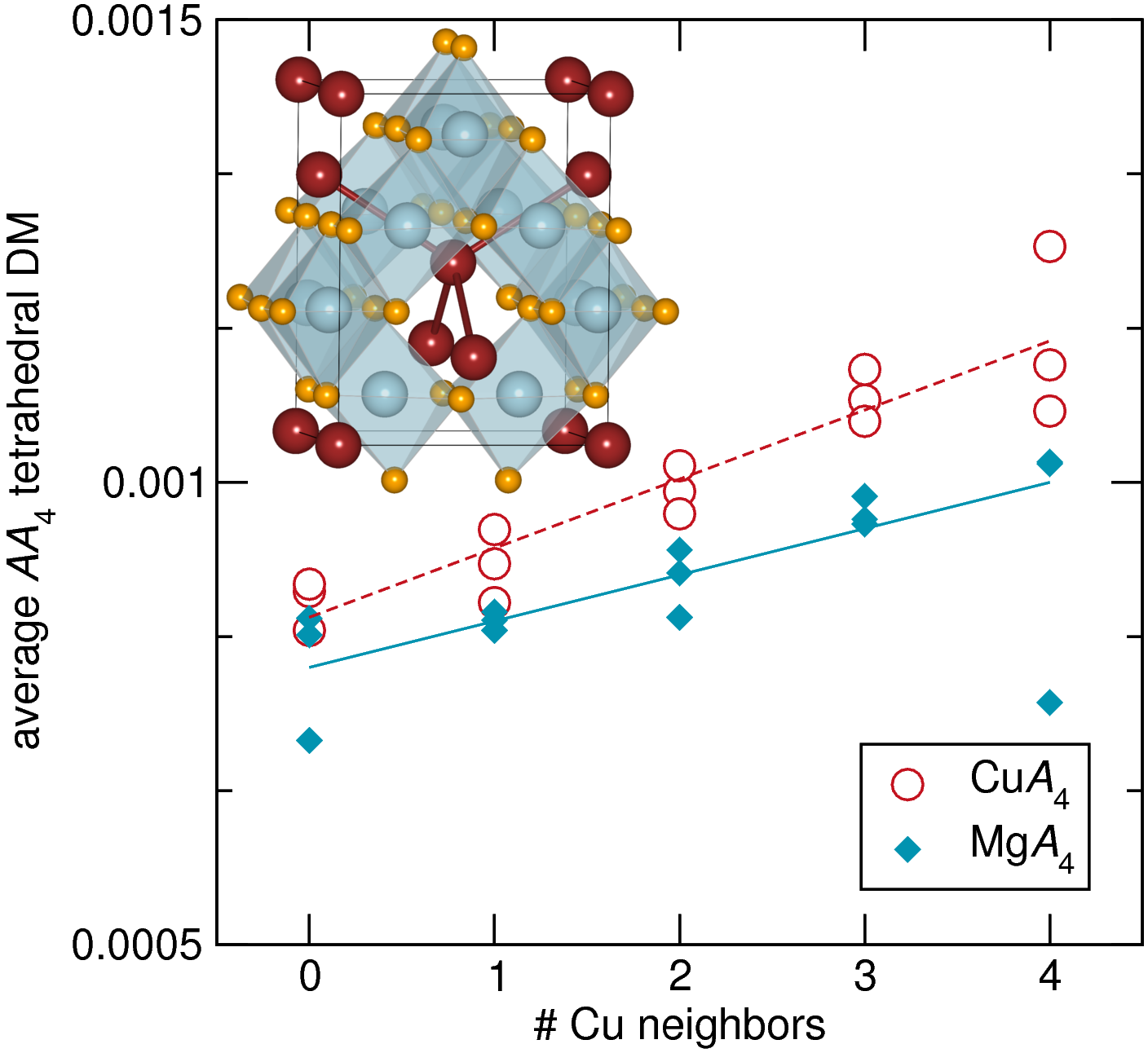}\\
\caption{(Color online) CSM distributions of $AA_4$ tetrahedra (connected
by long red bonds in inset) as a function of number of neighbors that
are JT-active Cu$^{2+}$. Distinct distributions are shown for Cu$A_4$ (circles) and
Mg$A_4$ (diamonds). Linear guides to the eye for each distribution show
increasing tetrahedral distortion with number of Cu neighbors, and overall
more distorted tetrahedra when tetrahedra are Cu-centered.
}
\label{fig:aa4-csm}
\end{figure}

Short-range interactions between JT-active cations are evident in the
$A$--$A$ correlations in Fig.\,\ref{fig:aa4-csm}. Each $A$ cation has four
$A$ nearest neighbors 3.59~\AA\ away, arranged in a diamond
lattice (inset in Fig.\,\ref{fig:aa4-csm}). The four nearest neighbors create a
large tetrahedron around the central cation. The tetrahedral DM for these
$AA_4$ are plotted for each central
cation (Mg or Cu) as a function of the number of Cu nearest neighbors, from 0 to 4.
Multiple RMC simulations show an upward 
trend indicating more distortion as the number of nearby Cu increases. The
separation between the two lines implies that Cu-centered Cu$A_4$ have
higher tetrahedral DM (more distortion) than Mg-centered Mg$A_4$. These
short-range correlations are indicative of strain coupling between
adjacent $A$-site JT distortions.

%% So too may be the addition of X-ray total scattering data in a combined fit.

%% One possibility would be to investigate the solid solution of CoCr$_2$O$_4$--\cco, which
%% has the Co$^{2+}$ $3d^7$, or CoCr$_2$O$_4$--NiCr$_2$O$_4$. Co$^{2+}$ is the only JT-inactive
%% divalent cation with a neutron scattering cross section distinctly different from
%% Cu, which is almost the same as Zn. 

%% somehow, the 20\% still manages a cooperative JT distortion even though there should
%% be very few Cu-Cu neighbors.

\subsection{Magnetic properties}

% CCO Tc = 135 or 133 K, MCO AFM Below 16 K

\mco\ and \cco\ have markedly different magnetic behavior due to the addition
of unpaired spins in Cu$^{2+}$ and the accompanying JT distortion. \cco\ is a
hard ferrimagnet with $T_C$ = 135 K,\cite{walter_magnetische_1967}
 while \mco\ undergoes complex antiferromagnetic
ordering below
$T_N$ = 16~K.\cite{shaked_magnetic_1970,ortega-san-martin_low_2008}
Macroscopic composites of a ferromagnet and antiferromagnet
would result in a traditional exchange biased
material, with an enhanced coercive field $H_C$ and an exchange bias field $H_E$,
manifested as a shift of the hysteresis loop in the
$-H$ direction.\,\cite{meiklejohn_new_1956,berkowitz_exchange_1999,nogus_exchange_2005}

\begin{figure}
\centering\includegraphics[width=0.9\columnwidth]{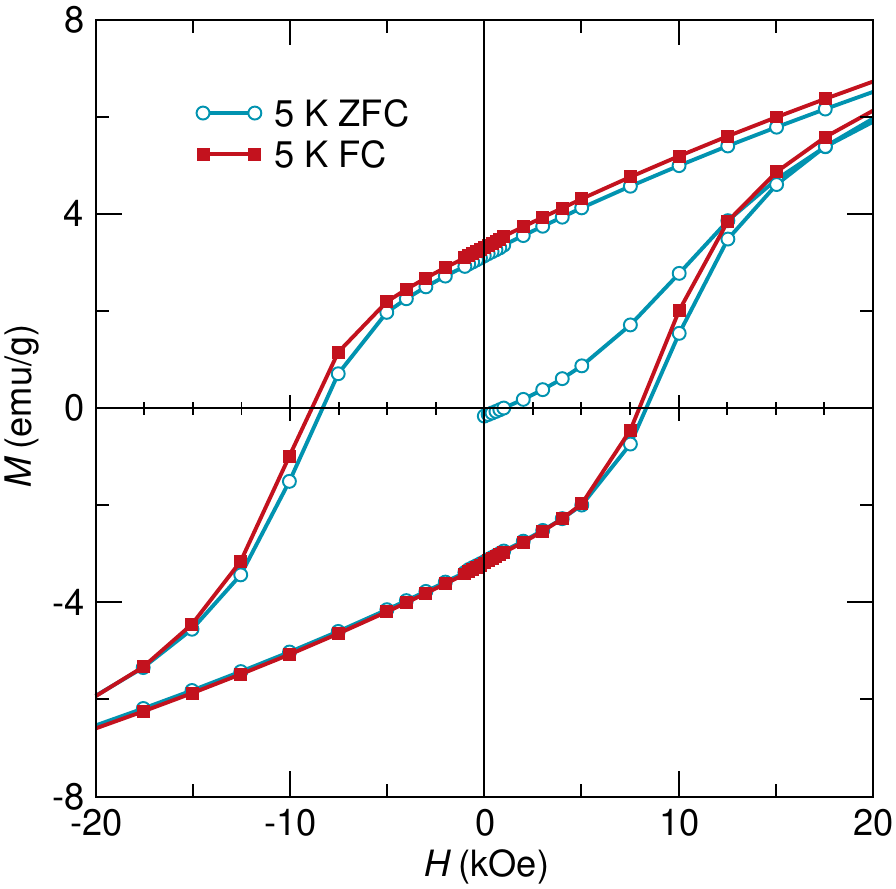}\\
\caption{(Color online) Magnetic hysteresis of \mcco\ with $x$ = 0.43 at
$T$ = 5~K after zero-field cooling (ZFC) and field cooling (FC) with
$H_{FC}$ = 5~T. The shift in the $-H$
direction when $M$ = 0 is the exchange bias field $H_E$.
}
\label{fig:hysteresis}
\end{figure}

\begin{figure}
\centering\includegraphics[width=0.9\columnwidth]{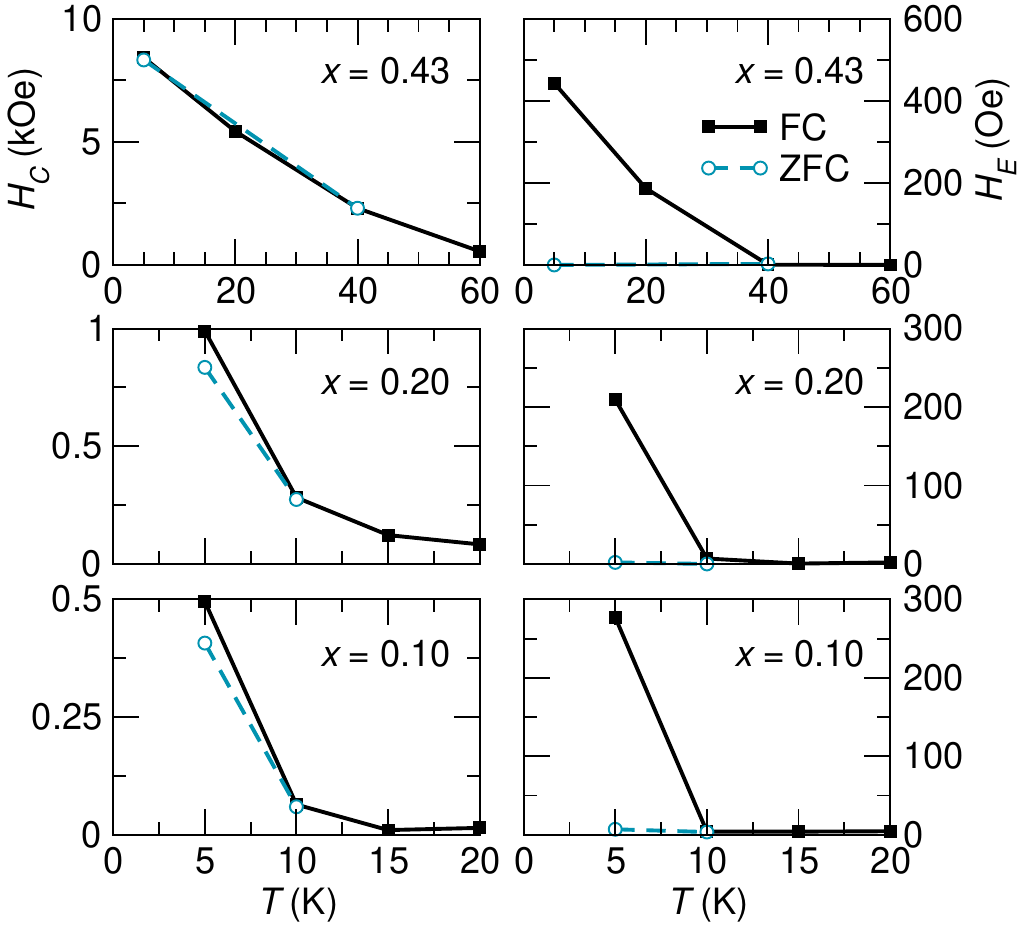}\\
\caption{(Color online) Coercive fields $H_C$ (left) and exchange bias fields $H_E$ (right)
for \mcco\ samples with $x$ = 0.43, 0.20, and 0.10 after 
ZFC and FC with $H_{FC}$ = 5~T. In all cases, ZFC results in a smaller $H_C$ and gives
$H_E$ = 0.}
\label{fig:exchangebias}
\end{figure}

The \mcco\ solid solution is a mixture on the atomic level, but
nevertheless exhibits the magnetic hallmarks of an exchange biased system.
The hysteresis loop of the $x$ = 0.43 sample in
Fig.\,\ref{fig:hysteresis} at $T$ = 5~K has $H_C = 8.4$~kOe.
Cooling with a field $H_{FC}$ = 50 kOe
broadens the hysteresis loop and shifts it in the $-H$ direction
by $H_E = 0.44$~kOe. This shift
signifies the preference for the ferrimagnet to align
along the field-cooling direction.
$H_C$ and $H_E$ decrease with the Cu$^{2+}$ concentration $x$,
as does the onset of magnetic ordering. These
trends are shown in Fig.\,\ref{fig:exchangebias}.
In all cases, field-cooling increases $H_C$ and results in the appearance
of a significant $H_E$. We find $H_E = 0$ in all samples after zero-field
cooling.

Exchange bias is traditionally manifested by pinning at the interface
between a ferro- or ferrimagnet and antiferromagnet after field cooling when
$T_C > T_N$. It can also arise from intrinsic disorder in a single-phase system,
or the presence of disordered spins from either a spin glass or uncompensated
surface spins on small particles
(which behave in a glassy manner
themselves).\,\cite{makhlouf_magnetic_1997,martinez_low_1998}
In the \mcco\ system, the magnetic behavior cannot be fully described by traditional
AFM-FM interplay between \mco\ and \cco\ because the onset of
$H_E$ when $x$ = 0.43 in Fig.\,\ref{fig:exchangebias}
occurs above $T$ = 20 K, which is above the N\'{e}el temperature of \mco. Therefore
the disordered solid solution contains some regions of heterogeneity which may 
behave as glassy moments or as some intermediate AFM phase. 
Either of these cases can produce exchange bias. \cite{ali_exchange_2007,yan_glassy_2008}

The correlations between $A$ cations seen in Fig.\,\ref{fig:aa4-csm}
indicate that the local structure varies with the composition of nearby
atoms. Magnetic interactions, in turn, will be affected by these local distortions.
A more detailed investigation of magnetic behavior, as was performed on
Zn$_x$Mn$_{3-x}$O$_4$, \cite{shoemaker_intrinsic_2009}
may help elucidate how the competing structural and magnetic interactions produce
exchange bias. The phenomenon
of JT-active cation clustering has been investigated by dilatometry, \cite{brabers_cation_1971}
but its effects on magnetism have not been explored in detail.

\section{Conclusions}

We find that the solid solution \mcco\ has a two-phase coexistence of cubic
and tetragonal phases at room temperature for $0.43 \leq x \leq 0.47$.
Tetragonality is induced by increasing JT activity in tetrahedrally
coordinated Cu$^{2+}$. The average structure descriptions from Rietveld refinement 
provide an adequate description of the structures across the range
of $x$ and temperature. This is corroborated by the magnetic behavior, which
indicates a disordered atomic mixture. The $x$ = 0.20 sample
is cooperatively JT distorted with orbital
ordering at 15 K despite 82\% of all Cu$^{2+}$ having zero or only
one Cu$^{2+}$ neighbors.

Least-squares PDF refinements achieve good fits using the models
from Rietveld refinement, implying that it might be difficult to improve
on models where CuO$_4$ and MgO$_4$ are equivalent. 
Still, bond valence calculations show that  RMC simulations produce better fits
to the data while retaining chemically reasonable bond distances.
The $A$O$_4$ tetrahedral distortion increases with $x$ as judged by
CSM. The absence of distinct CuO$_4$ local distortions in the low-$x$ CSM
histograms does not prohibit their existence. When $x$ = 0.43,
tetrahedral DM indicate distinct coordination of Cu$^{2+}$ versus
Mg$^{2+}$. Furthermore, cation-cation interactions probed by $AA_4$ CSM
indicate that local clustering of Cu leads to increased JT distortion.
This technique resolves cation-dependent JT distortions (even when they
are incoherent) in materials where contrast between cations exists.
The presence of magnetic exchange bias implies that short-range
structural details are influencing the magnetic interactions, and 
more complex magnetic characterization may help describe these interactions.
%% By swapping the distinct coordination environments, we can recover the
%% details in the PDF drive the RMC fits to produce this model with
%% chemically distinct environments on the spinel $A$ site.

\section{Acknowledgments}

We thank Maosheng Miao, Katharine Page, Graham King, Anna Llobet, 
and Thomas Proffen for
helpful discussions. Details of $AA_4$ correlations were investigated
at the suggestion of a referee. This work has benefited from the use
of NPDF at the Lujan Center at Los Alamos Neutron Science
Center, funded by DOE Office of Basic Energy Sciences. 
Los Alamos National Laboratory is operated
by Los Alamos National Security LLC under DOE Contract DE-AC52-06NA25396.
This work was supported by the Institute for Multiscale Materials Studies
%% , the donors of the American Chemical Society Petroleum Research Fund,
and the 
National Science Foundation (DMR 0449354).  
The MRL Central Facilities are supported by the MRSEC Program of the NSF
(DMR 05-20415); a member of the NSF-funded Materials
Research Facilities Network (www.mrfn.org).

\bibliography{mcco}

\clearpage

\end{document}